\documentstyle[preprint,prd,aps]{revtex}

\begin{document}
\draft
\title{The Mass Effect in Quarkonium Decay}
\author{J.P. Ma}
\address{Institute of Theoretical Physics,\\
Academia Sinica, \\
P.O.Box 2735, Beijing 100080, China\\
e-mail: majp@itp.ac.cn}
\maketitle

\begin{abstract}
In the NRQCD factorization for productions and decays of a quakonium the
difference between the quarkonium mass and the twice of the heavy quark mass
is neglected at the leading order of the small velocity expansion. The
effect of this difference is included in the relativistic correction. We
attempt to define the difference in terms of NRQCD matrix elements and to
separate the effect of the difference in several decays. It turns out that
the total relativistic correcttion can be estimated by the difference.
Numerically the correction will enhance the decay widths considered here.%
\end{abstract}

\preprint{\ \vbox{
\halign{&##\hfil\cr
       AS-ITP-99-07\cr\cr}} }

\baselineskip=15pt

\vspace{-5mm}

\vskip20pt

\vfill \eject
\narrowtext

Quarkonium systems generally are thought as simpler than light hadrons and
it may be easier to handle them in the framework of QCD. However, only
recently we have been able to treat their decays and productions rigorously,
based on a factorization with non-relativistic QCD(NRQCD)\cite{BBL}, where
the effect of short distance is handled with perturbative QCD and the effect
of long distance is parameterized with NRQCD matrix elements. The
factorization is performed by utilizing the fact that a heavy quark moves
with a small velocity $v$ in the quarkonium rest frame and an expansion in $%
v $ can be employed. In this factorization, a quarkonium decay, e.g., like $%
J/\Psi $, can be imagined at the leading order of $v$ as the following: The $%
c$- and $\bar{c}$- quark in $J/\Psi $ has certain probability to be freed at
the same space point and this $c\bar{c}$ pair decays subsequently. The
probability has a nonperturbative nature and is at order of $v^0$, while the
decay of the $c\bar{c}$ pair can be treated with perturbative QCD. Various
effects at higher order of $v$ can be taken into account, e.g., relativistic
effect, and the effect of that the freed $c\bar{c}$ pair does not possess
the same quantum numbers as those of $J/\Psi $. In order to make the
factorization consistently the quarkonium mass is forced to be approximated
as twice of the heavy quark mass. In general the quarkonium mass is not the
same as twice of the heavy quark mass, the difference, we call it as the
mass difference, is at order of $v^2$ and is made by nonperturbative
physics, which binds the heavy-, antiheavy quark and other possible light
dynamical freedoms into a bound state. This difference makes the phase space
of a quarkonium decay differently than that of a heavy quark pair decay and
introduces a correction at order of $v^2$. The situation here is relatively
similar as this in deeply inelastic scattering, where the mass of the
initial hadron introduces the target mass effect\cite{Na}. This effect can
be thought as a kinematical effect at higher-twist, the other higher twist
effects are dynamical, which come from the fact that partons inside a target
can have a transverse momentum and can be off-shell, $\cdots $ etc., these
together are corrections to the parton model. In a quarkonium decay, like $%
J/\Psi $-decay, the correction from the next-to-leading order of $v$ is
characterized by a single matrix element in NRQCD and both kinematic- and
dynamical effects are included in the correction, as shown below explicitly
in considered processes. However, the kinematic effect can be separated from
the high-order corrections, and it can be interpreted in term of the mass
difference. A decay width obtained with and without the separation should be
at the same accuracy at the considered orders. This fact leads to that the
effect at the next-leading order of $v^2$ can be interpreted totally by the
mass difference in the decays considered here. In this work we will first
try to define the mass difference in terms of NRQCD matrix elements and then
calculate the correction from the next-to-leading order with and without the
separation.

We start with the energy momentum tensor $T_{\mu \nu }$ of QCD, a nice
discussion about properties of the tensor and interesting references can be
found in\cite{Ji1}. For our purpose we consider the full QCD containing
gluons, $N_f$ flavors of light quarks and one flavor of heavy quark $Q$. The
tensor then can be written as 
\begin{equation}
T^{\mu \nu }(x)=T_Q^{\mu \nu }(x)+T_G^{\mu \nu }(x)+T_q^{\mu \nu }(x)
\end{equation}
where $T_Q^{\mu \nu }$ is the contribution from the heavy quark, $T_G^{\mu
\nu }$ is from gluons and $T_q^{\mu \nu }$ is from light flavors. For a
state of a hadron with momentum $p$ one has by definition: 
\begin{equation}
\frac{\langle p|\int d^3xT^{0\mu }(x)|p\rangle }{\langle p|p\rangle }=p^\mu .
\end{equation}
If the hadron is at rest, the zeroth component in the right side of Eq.(2)
is the mass $M$ of the hadron. The energy momentum tensor can be decomposed
into a trace part and a traceless part, 
\begin{eqnarray}
T^{\mu \nu }(x) &=&\bar{T}^{\mu \nu }(x)+\hat{T}^{\mu \nu }(x), \\
g_{\mu \nu }\bar{T}^{\mu \nu }(x) &=&0.
\end{eqnarray}
From the space-time covariance one can derive 
\begin{eqnarray}
p^0\langle p|T_{\mu \nu }|p\rangle &=&p_\mu p_\nu , \\
p^0\langle p|\bar{T}_{\mu \nu }|p\rangle &=&p_\mu p_\nu -\frac 14g_{\mu \nu
}M^2, \\
p^0\langle p|\hat{T}_{\mu \nu }|p\rangle &=&\frac 14g_{\mu \nu }M^2
\end{eqnarray}
where the factor $p^0$ appears because we take the normalization of a state
as $\langle p|p^{\prime }\rangle =(2\pi )^3\delta ^3({\bf p}-{\bf p^{\prime }%
})$. To interpret the mass difference we match the heavy quark fields of
full QCD into NRQCD fields. We perform this matching at the tree-level and
for consistency we neglect the contributions from the trace-anomaly and the
anomalous dimension of quark masses. The heavy quark contribution $T_Q^{\mu
\nu }$ in the full QCD reads: 
\begin{eqnarray}
T_Q^{\mu \nu } &=&\bar{T}_Q^{\mu \nu }+\hat{T}_Q^{\mu \nu }, \\
\bar{T}_Q^{\mu \nu } &=&\frac 12\{\bar{Q}(iD^\mu \gamma ^\nu +iD^\nu \gamma
^\mu )Q-\frac 12g^{\mu \nu }\bar{Q}i\gamma \cdot DQ\}, \\
\hat{T}_Q^{\mu \nu } &=&\frac 14g^{\mu \nu }m_Q\bar{Q}Q.
\end{eqnarray}
In the above $Q$ is the dirac field for the heavy quark $Q$, $D^\mu $ is the
covariant derivative. The last term with $g^{\mu \nu }$ in $\bar{T}_Q^{\mu
\nu }$ can be simplified with the equation of motion. We take the state $%
|p\rangle $ as a quarkonium state $|H\rangle $ and match the matrix element
of $\hat{T}_Q^{\mu \nu }$ into NRQCD matrix elements, we obtain: 
\begin{eqnarray}
M_H &=&2m_Q-\langle H|T_K|H\rangle +\sum_fm_f\langle H|\bar{q}_fq_f|H\rangle
+O(v^4)+O(\alpha _s) \\
T_K &=&-\frac 1{2m_Q}\psi _Q^{\dagger }{\bf D}^2\psi _Q+\frac 1{2m_Q}\chi
_Q^{\dagger }{\bf D}^2\chi _Q,
\end{eqnarray}
where $T_K$ is an operator defined in NRQCD, $\psi _Q$ and $\chi _Q$ are the
quark and antiquark fields in NRQCD respectively, $M_Q$ is the pole mass of
the heavy quark. The operator $T_K$ measures the sum of kinetic energies of
a heavy and antiheavy quark. For $H$ with a nonzero spin the average of the
spin is implied. The summation is over all light flavors. The matrix
elements $\langle H|\bar{q}_fq_f|H\rangle $ are at least at the order of $%
v^2 $. This can be estimated by taking a $Q\bar{Q}$ pair instead of the
state $H$, because the operator $\bar{q}_fq_f$ is a color-singlet, the
matrix element is nonzero provided that at least two gluons must be
exchanged between the operator and the $Q\bar{Q}$ state. A gluon coupled to $%
Q$ or $\bar{Q}$ gives a factor $v$. The contribution from this operator to $%
M_H$ is suppressed by a factor $m_f/m_Q$, relatively to the contribution
from $T_K$, can be neglected for charmonium as the mass ratio is small
enough. For bottonium system, because charm quarks are included in the full
QCD, neglecting this contribution from charm quark may be problematic,
because the mass ratio is not so small. However the matrix element $\langle
H|\bar{c}c|H\rangle $, where $c$ and $\bar{c}$ is the dirac field for charm
quark, is suppressed by $v^2\Lambda _{{\rm QCD}}^2/m_c^2$ at least, hence
the contribution from this operator suppressed by $\Lambda _{{\rm QCD}%
}^2/m_bm_c$ relatively to that from $T_K$. Giving the fact that $\Lambda _{%
{\rm QCD}}$ as a characteristic scale for long-distance physics is at order
of several hundred MeV, this contribution can be neglected too.

Neglecting the contributions proportional to $m_f$ and other higher order
effects we obtain 
\begin{equation}
M_H = 2m_Q -\langle H \vert T_K \vert H \rangle .
\end{equation}
Taking this result we obtain another relation in the quarkonium rest frame
within our approximation: 
\begin{equation}
2\langle H\vert T_K \vert H\rangle =-\langle H \vert T_G^{00} +T_q ^{00}
\vert H \rangle .
\end{equation}
This relation may be regarded as a field theory version of the virial
theorem of quantum mechanics, which relates the expectation value of the
kinetic energy to the expectation value of the potential energy. The above
equations hold only at the tree-level approximation in field theory. If one
takes the contributions from the trace-anomaly, the contributions from the
anomalous dimension of quark masses and those from the matching beyond the
tree-level, into account, these relations become more complicated. For
higher exited states of quarkonia the relation needs to be corrected by
higher orders of $v^2$ as $v^2$ becomes larger for these states. The matrix
element $\langle H\vert T_K \vert H\rangle $ is positive and it indicates
that $M_H<2m_Q$.

If the mass of a quakonium and the pole mass of the heavy quark are known,
one may determines how large the matrix element $\langle H\vert T_K \vert
H\rangle $ or the binding energy is. We take $\Upsilon$ and $J/\Psi$ as
examples. The pole mass of b-quark is determined precisely from a study of
lattice QCD\cite{mb}, whose result is $m_b=5.0$GeV. For charm quark we take
the pole mass determined from the D meson semileptonic decay\cite{mc}, the
value is $m_c=1.65$GeV. We obtain: 
\begin{eqnarray}
\langle \Upsilon \vert T_K \ \vert \Upsilon \rangle &\approx & 0.54{\rm GeV},
\\
\langle J/\Psi\vert T_K \vert J/\Psi \rangle &\approx & 0.20{\rm GeV}.
\end{eqnarray}
With these values one can also determine the velocity of b- or c-quark in $%
\Upsilon$ or $J/\Psi$ respectively, by identifying $\langle H\vert T_K \vert
H\rangle =m_Q v^2$. We obtain $v^2\approx 0.1$ for b-quark and $v^2\approx
0.12$ for c-quark. The velocity for c-quark is smaller than that from usual
estimation. For c-quark it should be noted that a precise determination of
the pole mass is still lacking and the higher order effect in Eq.(13) may be
significant.

Now we consider the decay $J/\Psi \rightarrow \ell ^{+}\ell ^{-}$. Starting
with the relevant $S$-operator we obtain the $S$-matrix element 
\begin{eqnarray}
\langle \ell ^{+}(k_1),\ell ^{-}(k_2)|S|J/\Psi (P)\rangle &=&-ie^2Q_c(2\pi
)^4\delta ^4(P-k_1-k_2)  \nonumber \\
&\cdot &\int \frac{d^4q}{(2\pi )^4}\bar{u}(k_2)\gamma _\mu v(k_2)\frac 1{q^2}%
{\rm Tr}(\gamma ^\mu \Gamma (q,P))  \nonumber \\
\Gamma_{ij} (q,P) &=&\int d^4xe^{iq\cdot x}\langle 0|\bar{c}%
(x)_jc(x)_i|J/\Psi \rangle
\end{eqnarray}
To perform the NRQCD factorization we match the nonperturbative object $%
\Gamma (q,P)$ into NRQCD matrix elements by the small velocity expansion. Up
to the order of $v^2$ it reads: 
\begin{eqnarray}
\Gamma_{ij} (q,P) &=&(2\pi )^4\delta ^4(q-p_0)[-\frac 12(P_{+}\gamma
^lP_{-})_{ij}\langle 0|\chi _c^{\dagger }\sigma ^l\psi _c|J/\Psi \rangle 
\nonumber \\
&+&\frac 12(P_{-}\gamma ^lP_{+})_{ij}\frac 1{12m_c^2}\langle 0|({\bf D^2}%
\chi _c)^{\dagger }\sigma ^l\psi _c|J/\Psi \rangle  \nonumber \\
&-&\frac 12(P_{+}\gamma ^lP_{-})_{ij}\frac 1{4m_c^2}\langle 0|({\bf D^2}\chi
_c)^{\dagger }\sigma ^l\psi _c|J/\Psi \rangle ]  \nonumber \\
&-&\frac \partial {\partial q^0}(2\pi )^4\delta ^4(q-p_0)(P_{+}\gamma
^lP_{-})_{ij}\frac 1{m_c}\langle 0|({\bf D^2}\chi _c)^{\dagger }\sigma
^l\psi _c|J/\Psi \rangle
\end{eqnarray}
where $p_0^\mu =(2m_c,0,0,0)$, $P_{\pm }=(1\pm \gamma ^0)/2$. The first term
is at order of $v^0$, the remaining terms are at order of $v^2$. The
relativistic correction is not only from these terms but also from $P^\mu
=(M_{J/\Psi },0,0,0)$ in the $S$-matrix element. For the terms at $v^2$ in $%
\Gamma (q,P)$ it can be replaced by $P=p_0$ because we neglect the effect at
higher orders. For the first term in $\Gamma (q,P)$ we can write the $S$%
-matrix element with translational invariance as: 
\begin{eqnarray}
\langle \ell ^{+}(k_1),\ell ^{-}(k_2)|S|J/\Psi (P )\rangle &=&ie^2Q_c\frac 1{%
4m_c^2}\bar{u}(k_2)\gamma ^iv(k_2)(2\pi )^4\delta ^4(P-k_1-k_2)\langle
0|\chi _c^{\dagger }\sigma ^i\psi _c|J/\Psi \rangle  \nonumber \\
&+&\cdots \cdots  \nonumber \\
&=&ie^2Q_c\frac 1{4m_c^2}\bar{u}(k_2)\gamma ^iv(k_2)\int dx^4e^{ix\cdot
(k_1+k_2-P)}\langle 0|\chi _c^{\dagger }\sigma ^i\psi _c|J/\Psi \rangle 
\nonumber \\
&+&\cdots \cdots  \nonumber \\
&=&ie^2Q_c\frac 1{4m_c^2}\bar{u}(k_2)\gamma ^iv(k_2)\int dx^4e^{ix\cdot
(k_1+k_2-p_0)}\langle 0|\chi _c^{\dagger }(x)\sigma ^i\psi _c(x)|J/\Psi
\rangle  \nonumber \\
&+&\cdots \cdots
\end{eqnarray}
where $\cdots \cdots $ stands for the contributions from the terms at $v^2$
in $\Gamma (q,P)$. The $x$-dependence in the NRQCD matrix element can be
expanded. This expansion will lead to a tower of operators with $n$%
-derivative with $x^0$. Similar cases in quarkonium productions are
considered\cite{BRW}, where one can add the effect due to these operators to
cross-sections at the leading order of $v$ and it results in that the
hadronic phase-space is recovered. Here we start with the hadronic
phase-space and try to approximate it with the partonic-phase space. After
the expansion we obtain the $S$-matrix element up to order of $v^2$: 
\begin{eqnarray}
\langle \ell ^{+}(k_1),\ell ^{-}(k_2)|S|J/\Psi (P )\rangle &=&ie^2Q_c\bar{u}%
(k_2)\gamma ^iv(k_2)  \nonumber \\
&\cdot &\big\{(2\pi )^4\delta ^4(p_0-k_1-k_2)(\frac 1{4m_c^2}\langle 0|\chi
_c^{\dagger }\sigma ^i\psi _c|J/\Psi \rangle  \nonumber \\
&+&\frac 7{24m_c^4}\langle 0|({\bf D^2}\chi _c)^{\dagger }\sigma ^i\psi
_c|J/\Psi \rangle )  \nonumber \\
&+&\frac 1{4m_c^3}\frac \partial {\partial k_1^0}(2\pi )^4\delta
^4(p_0-k_1-k_2)\langle 0|({\bf D^2}\chi _c)^{\dagger }\sigma ^i\psi
_c|J/\Psi \rangle \big\}
\end{eqnarray}
With the above $S$-matrix element we obtain the decay width 
\begin{eqnarray}
\Gamma (J/\Psi \rightarrow \ell ^{+}\ell ^{-}) &=&\alpha ^2Q_c^2\frac{2\pi }3%
\frac 1{m_c^2}\big [|(\langle 0|\chi _c^{\dagger }{\bf \sigma }\psi
_c|J/\Psi \rangle |^2  \nonumber \\
&+&\frac 2{3m_c^2}(\langle J/\Psi |\psi _c^{\dagger }{\bf \sigma }\chi
_c|0\rangle \cdot \langle 0|({\bf D^2}\chi _c)^{\dagger }{\bf \sigma }\psi
_c|J/\Psi \rangle +h.c.)\big ]+O(v^4)
\end{eqnarray}
where the spin average for $J/\Psi $ is implicit in the products of the
matrix elements. This result agrees with that derived by the matching
procedure in \cite{BBL}. The relativistic correction is represented by the
product of matrix elements: 
\begin{equation}
a=\langle J/\Psi |\psi _c^{\dagger }{\bf \sigma }\chi _c|0\rangle \cdot
\langle 0|({\bf D^2}\chi _c)^{\dagger }{\bf \sigma }\psi _c|J/\Psi \rangle
+h.c..
\end{equation}
The decay width depends on $M_{J/\Psi}$ only implicitly through the matrix
elements, while originally there is an explicit dependence on $M_{J/\Psi}$
in the term detailed in Eq.(19). However this term can be treated by the
direct expansion of the $\delta $-function 
\begin{eqnarray}
\delta (M_{J/\Psi }-k_1^0-k_2^0) &=&\delta (2m_c-k_1^0-k_2^0)-\Delta
M_{J/\Psi }\frac \partial {\partial k_1^0}\delta (2m_c-k_1^0-k_2^0)+{\cal O}%
(v^4)  \nonumber \\
\Delta M_{J/\Psi } &=&M_{J/\Psi }-2m_c
\end{eqnarray}
where $\Delta M_{J/\Psi }$ is at order of $v^2$ as shown before. With this
expansion and $\Gamma (q,P)$ given in Eq.(18) the decay width is 
\begin{eqnarray}
\Gamma (J/\Psi \rightarrow \ell ^{+}\ell ^{-}) &=&\alpha ^2Q_c^2\frac{2\pi }3%
\frac 1{m_c^2}\big [|\langle 0|\chi _c^{\dagger }{\bf \sigma }\psi _c|J/\Psi
\rangle |^2(1+\frac{\Delta M_{J/\Psi }}{m_c})  \nonumber \\
&+&\frac 7{6m_c^2}a\big ]+{\cal O}(v^4).
\end{eqnarray}
Both results should be equivalent up to order of $v^2$, one can hence find a
relation between $a$ and $\Delta _{J/\Psi }$ and rewrites the decay width as 
\begin{equation}
\Gamma (J/\Psi \rightarrow \ell ^{+}\ell ^{-})=\alpha ^2Q_c^2\frac{2\pi }3%
\frac 1{m_c^2}|(\langle 0|\chi _c^{\dagger }{\bf \sigma }\psi _c|J/\Psi
\rangle |^2(1-\frac{4\Delta M_{J/\Psi }}{3m_c})+{\cal O}(v^4).
\end{equation}

As discussed before, the quantity $M_{J/\Psi }$ is negative, hence the
relativistic correction seems to enhance the decay width. This is in
contrast to the model calculation where one uses a wave-function computed in
nonrelativistic potential models. With the wave-function the parameter $a$
can be estimated as 
\begin{equation}
a=\frac 1{2\pi }R^{*}(r){\bf \bigtriangledown}^2R(r)|_{r=0}+h.c.
\end{equation}
where $R(r)$ is the radial wave function. From above $a$ is negative, the
decay width therefor is reduced. But as discussed in \cite{BBL}, this
parameter estimated in this way is not well defined because it is divergent
if the potential is Coulombic at short distances. Certain subtraction is
need to make it finite. It is unclear how to do the subtraction in a model
calculation to make the estimation meaningful for the renormalized matrix
element in $a$. In model calculations certain regularization is used to
obtain a finite relativistic correction, like in \cite{CHL}.

In rewriting the decay width we used the relation between $a$ and $\Delta
M_{J/\Psi }$. It should be kept in mind that this relation is not proved in
general and it applies only in the detailed cases considered here. Similar
analysis can be done for $\eta _c\rightarrow \gamma \gamma $ and the result
is that the kinematic effect does not appear at the next-to-leading order of 
$v$.

The other process to be considered in this work is $J/\Psi \rightarrow {\rm %
light\ hadrons}$. It should be noted that for processes involving strong
interaction NRQCD factorization is in general performed for squares of
amplitudes. However, the vacuum saturation is a good approximation in NRQCD,
this enables us to work at the level of amplitude. With the vacuum
saturation and at leading order of $\alpha _s$ the process to be considered
is $J/\Psi \rightarrow {\rm 3\ Gluons}$. The $S$-matrix element can be
written: 
\begin{eqnarray}
\langle 3G|S|J/\Psi (P)\rangle  &=&i(2\pi )^4\delta ^4(P-k_1-k_2-k_3)\int 
\frac{d^4q}{(2\pi )^4}{\rm Tr}\{A(k_1,k_2,k_3,q)\Gamma ^h(q,P)\}, \\
\Gamma _{ij}^h(q) &=&\int d^4xe^{iq\cdot x}\langle 0|\bar{c}%
_j(x)c_i(0)|J/\Psi \rangle , \\
A(k_1,k_2,k_3,q) &=&-ig_s^3T^{a_1}T^{a_2}T^{a_3}\frac 1{(k_1-q)^2-m_c^2)}%
\frac 1{(k_1+k_2-q)^2-m_c^2)}  \nonumber \\
&\cdot &\gamma \cdot \varepsilon _1^{a_1}(\gamma \cdot (k_1-q)+m_c)\gamma
\cdot \varepsilon _2^{a_2}(\gamma \cdot (k_1+k_2-q)+m_c)\gamma \cdot
\varepsilon _3^{a_3}  \nonumber \\
&+&{\rm Permutation\ of\ 1,2,3}.
\end{eqnarray}
The indices 1,2,3 are labels for the 3 gluons. To calculate the decay width
up to $v^2$ we match the nonperturbative object $\Gamma ^h(q)$ into NRQCD
matrix element and only keep the contributions up to order of $v^2$ and take
the contribution at $v^2$ from the $\delta $-function for the energy
conservation into account. In the calculation of the effect at $v^2$ one
should use physical spin-sums for gluons and be careful with the integration
over the 3-body phase-space. The detailed calculation is tedious but
straightforward. We obtain: 
\begin{eqnarray}
\Gamma (J/\Psi \rightarrow {\rm light\ hadrons}) &=&\frac{20}{243m_c^2}%
\alpha _s^3\big\{|\langle 0|\chi _c^{\dagger }{\bf \sigma }\psi _c|J/\Psi
\rangle |^2(\pi ^2-9)  \nonumber \\
&+&(\frac{101}{192}\pi ^2-\frac 72)\cdot \frac 1{m_c^2}a\big\}+{\cal O}(v^4).
\end{eqnarray}
Similarly as for the leptonic decay we can directly expand the $\delta $%
-function for the energy conservation and obtain the decay width as 
\begin{eqnarray}
\Gamma (J/\Psi \rightarrow {\rm light\ hadrons}) &=&\frac{20}{243m_c^2}%
\alpha _s^3\big\{|\langle 0|\chi _c^{\dagger }{\bf \sigma }\psi _c|J/\Psi
\rangle |^2(\pi ^2-9+\frac{352-33\pi ^2}{32}\frac{\Delta M_{J/\Psi }}{m_c}) 
\nonumber \\
&+&\frac{192+\pi ^2}{96}\cdot \frac 1{m_c^2}a\big\}+{\cal O}(v^4).
\end{eqnarray}
Both widths are correct up to order of $v^2$, we can hence rewrite the width
as: 
\begin{eqnarray}
\Gamma (J/\Psi \rightarrow {\rm light\ hadrons}) &=&\frac{20}{243m_c^2}%
\alpha _s^3|\langle 0|\chi _c^{\dagger }{\bf \sigma }\psi _c|J/\Psi \rangle
|^2\big [\pi ^2-9-(\frac{101}{96}\pi ^2-7)\frac{\Delta M_{J/\Psi }}{m_c}\big
]  \nonumber \\
&+&{\cal O}(v^4).
\end{eqnarray}
If we take $m_c=1.65$GeV as before, the width for the decay into leptons and
into light hadrons is enhanced at the level of $16\%$ and of $48\%$
respectively.. The hadronic width receives a substantial correction. From
Eq.(32) and Eq.(25) one can also obtain the corresponding widths for $%
\Upsilon $. The same level of the enhancement is also found for $\Upsilon
\rightarrow \ell ^{+}\ell ^{-}$ and for $\Upsilon \rightarrow {\rm light\
hadrons}$. To compare with experimental data we build the ratio: 
\begin{eqnarray}
r_b &=&\frac{\Gamma (\Upsilon \rightarrow {\rm light\ hadrons})}{\Gamma
(\Upsilon \rightarrow \ell ^{+}\ell ^{-})}  \nonumber \\
&=&5040.0\alpha _s^3(m_b)(1.276-0.081\alpha _s(m_b))
\end{eqnarray}
where we have taken $M_\Upsilon =9.460$GeV and the pole mass $m_b=5.0$GeV.
We have also added the one-loop QCD correction in each decay width\cite
{O1,ML}. In this ratio the one-loop correction is small, while the
correction in each decay width is large. With the experimental value $%
r_b=37.04$ we can estimate $\alpha _s(mb)$ as 
\begin{equation}
\alpha _s(m_b)=0.18.
\end{equation}
which is close to the value of $\alpha _s$ by running down from the scale $%
\mu =M_Z$. However, the relativistic correction is not significant in the
determination of $\alpha _s$. Similar estimation can be done for $J/\Psi $,
the determined $\alpha _s(m_c=1.6{\rm GeV})$ is 0.195. It is smaller than
the expected. This indicates that effects in higher orders of $v$ may be
still significant. It also should be noted that the pole mass $m_c$ is not
known precisely as $m_b$.

To summarize: In this work we tried to define the difference between a
quarkonium mass and the twice of the heavy quark mass in terms of NRQCD
matrix elements. We separated the effect introduced by the difference in
relativistic corrections for several decays of quarkonium and find that the
relativistic correction is determined by the difference. With the
determination decay widths considered here are enhanced by the relativistic
correction.

\vskip20pt \noindent
{\bf Acknowledgment:} The author would like to thank the unknown referee for
pointing out that the starting point in the previous version of the paper is
not correct. This work is supported by the ``Hundred Yonng Scientist
Program" of Sinica Academia of P.R.China.

\vskip15pt

\vfil\eject


\begin{references}
\bibitem{BBL}  G.T Bodwin, E. Braaten, and, G.P Lepage, Phys. Rev. D {\bf 51}
1125 (1995).

\bibitem{Na}  O. Nachtmann, Nucl. Phys. B63 (1973) 237, ibid B78 (1974) 455

\bibitem{Ji1}  X. Ji, Phys. Rev. D52 (1995) 271

\bibitem{mc}  V. Chernyak, Nucl. Phys. B457 (1995) 96

\bibitem{mb}  C.T.H. Davies et al, Phys. Rev. Lett. 73 (1994) 2654

\bibitem{BRW}  M. Beneke, I.Z. Rothstein and M.B. Wise, Phys. Lett. B408
(1997) 373

\bibitem{CHL}  K.T. Chao, H.W. Huang and Y.Q. Liu, Phys. Rev. D53 (1996) 221

\bibitem{O1}  R. Barbieri, R. Gatto, R. K\"{o}geler and Z. Kunszt, Phys.
Lett. B57 (1975) 455

W. Celmaster, Phys. Rev. D19 (1979) 1517

\bibitem{ML}  P. Mackenzie and G.P. Lepage, Phys. Rev. Lett. (1981) 1244
\end{references}
\end{document}